\documentstyle[preprint,aps]{revtex}
\begin{document}
\baselineskip=24ptplus.5ptminus.2pt
\vspace*{0.5 in}

\large
\begin{center}
{\bf Simultaneous Optical Model Analyses of Elastic Scattering, \\ 
Breakup, and Fusion Cross Section Data for the \\
$^{6}$He + $^{209}$Bi System at Near-Coulomb-Barrier Energies}
\end{center}
\normalsize
\vspace{0.5cm}

\begin{center}
B. T. Kim, W. Y. So, and S. W. Hong \\
{\it Department of Physics and Institute of Basic Science, \\
Sungkyunkwan University, Suwon 440-746, Korea}\\
\parskip 2ex
T. Udagawa \\
{\it Department of Physics, University of Texas, Austin, Texas 78712}
\end{center}

\vspace{0.5cm}
\begin{center}
{\bf Abstract}
\end{center}

\baselineskip=24ptplus.5ptminus.2pt
Based on an approach recently proposed by us, simultaneous $\chi^{2}$-analyses 
are performed for elastic scattering, direct reaction (DR) and fusion cross 
sections data for the $^{6}$He+$^{209}$Bi system at near-Coulomb-barrier 
energies to determine the parameters of the polarization potential 
consisting of DR and fusion parts.
We show that the data are well reproduced by the resultant potential, 
which also satisfies the proper dispersion relation. 
A discussion is given of the nature of the threshold anomaly seen 
in the potential.

\vspace{1.5em}
24.10.-i,~25.70.Jj
\pagebreak

%\documentstyle[preprint,aps]{revtex}
%\begin{document}
%\baselineskip=24pt plus .5pt minus .2pt

%\maketitle
%\begin{abstract}
%\end{abstract}

%\vspace{2ex}

%\pacs{24.10.-i,~25.70.Jj}

\narrowtext

 A great deal of effort has recently been focused on studies of the so-called 
threshold anomaly~\cite{maha,naga} (rapid energy variation in the strength of 
the optical potential) in heavy ion scattering induced, particularly, 
by very loosely bound projectiles such as $^{6}$He~\cite{agu1}, 
$^{6}$Li~\cite{tied,keel,maci}, and $^{9}$Be~\cite{mora,sig1}. 
The experimental results accumulated so far indicate that 
the imaginary part of the optical potential, $W(r;E)$, extracted
by analysis of the elastic scattering data,  does not show
such an anomaly as is observed in the potentials for
normal, tightly bound projectiles.
For tightly bound projectiles, $W(r;E)$ at around
the strong absorption radius $r=R_{sa}$ is found to decrease rapidly 
as the incident energy $E$ falls below the Coulomb-barrier energy $E_{c}$, 
and eventually vanishes at some {\it threshold} energy $E_{0}$. 
Contrary to this, for loosely bound projectiles $W(R_{sa};E)$
remains large at energies even below $E_{c}$~\cite{agu1,keel,maci,sig1}. 

 The reason for $W(R_{sa};E)$ being so large at low energies has been 
ascribed to the weak binding of the extra neutrons to the core nucleus, 
leading to breakup.
In fact, the breakup cross sections have been measured 
for these projectiles~\cite{agu2,kell,sig2},
confirming that they are indeed large, even larger than the fusion
cross sections at $E \sim E_{c}$. 
It was argued~\cite{maha} that since the energy dependence of 
the polarization potential due to the breakup must be weak, 
one might not be able to observe a noticeable energy variation 
in the resultant potential when the breakup cross section is larger than 
the fusion cross section as for loosely bound projectiles. 

 The threshold anomaly of $W(r;E)$ observed for tightly bound projectiles 
may be ascribed to the coupling of the elastic and fusion channels~\cite{uda1}. 
This is substantiated by the fact that the threshold energy $E_{0}$ of $W(r;E)$
({\it i.e.}, the energy where $W(r,E_{0})=0$) agrees very well with 
that of the fusion cross section $\sigma_{F}$, or more precisely 
the threshold energy of $S(E) \equiv \sqrt{E \sigma_{F}}$~\cite{stel}. 
It is thus natural that if the breakup cross section is larger than 
the fusion cross section, and if one is concerned only with the total $W(r;E)$, 
the rapid change in the fusion cross section and the anomaly would not 
show up clearly in the total $W(r;E)$.
                                                           
 Insight into this problem may be obtained if one decomposes the total $W(r;E)$ 
into the direct reaction (DR) and fusion parts, $W_{D}(r;E)$ and $W_{F}(r;E)$, 
respectively, and determines them separately~\cite{kim1}. 
The aim of the present study is to make such a determination 
of $W_D (r;E)$ and $W_F (r;E)$ by performing simultaneous $\chi^{2}$-analyses 
of elastic scattering, DR (breakup), and fusion cross section data. 
We take the $^{6}$He+$^{209}$Bi system, for which data are available 
not only for elastic scattering~\cite{agu1}, but also for breakup~\cite{agu2} 
and for fusion~\cite{kol1}. 
Following Ref.~\cite{agu1}, we identify the breakup cross section with 
the DR cross section. 
Optical model analyses of the elastic scattering and total reaction cross 
section data have already been presented in Refs.~\cite{agu1,mohr}. 
The present analysis is thus an extension of the previous studies.                            

 The optical potential $U$ we use has the following form;
\begin{equation}
U=U_{C}(r)-[V_{0}(r)+V(r;E)+iW(r;E)],
\end{equation}
where $U_{C}(r)$ is the Coulomb potential, whose radius parameter is fixed  
at a standard value of $r_{c}$=1.25~fm, and $V_{0}(r)$ is
the energy independent Hartree-Fock part of the potential, 
while $V(r;E)$ and $W(r;E)$ are, respectively, real and imaginary parts 
of the so-called polarization potential~\cite{sat1} 
that originates from couplings to reaction channels.  
$W(r;E)$ is assumed to have a volume-type fusion and 
a surface-derivative-type DR part. 
Explicitly, $V_{0}(r)$  and $W(r;E)$ are given, respectively, by
\begin{equation}
V_{0}(r)=V_{0}f(X_{0})
\end{equation}
and
\begin{equation}
W(r;E) = W_{F}(r;E)+W_{D}(r;E) =
W_{F}(E)f(X_{F})+4W_{D}(E)a_{D}\frac{df(X_{D})}{dR_{D}},
\vspace{2ex}
\end{equation}
where $f(X_{i})=[1+\mbox{exp}(X_{i})]^{-1}$, with $X_{i}=(r-R_{i})/a_{i}$
$({\it i}=0, \; D\; \mbox{and} \; F)$, is the usual Woods-Saxon function.
The real part of the polarization potential is also assumed to have 
DR and fusion parts; $V(r;E) = V_{F}(r;E)+V_{D}(r;E)$.
Each real part may be generated from the corresponding imaginary potential 
by using the dispersion relation~\cite{maha};
\begin {equation}
V_{i}(r;E)=V_{i}(r;E_{s}) + \frac {E-E_{s}}{\pi } \mbox{P} 
\int_{0}^{\infty} dE'
\frac {W_{i}(r;E')}{(E'-E_{s})(E'-E)},
\vspace{2ex}
\end {equation}
where P stands for the principal value and $V_{i}(r;E_{s})$ is the value
of the potential at a reference energy $E=E_{s}$. 
Later, we will use Eq.~(4) to generate the final real polarization potentials 
$V_{F}(r;E)$ and $V_{D}(r;E)$,  after $W_{F}(r;E)$ and 
$W_{D}(r;E)$ have been fixed from $\chi^{2}$-analyses.
For $V_{0}(r)$, we simply use the potential determined for the 
$\alpha$+$^{209}$Bi system at $E$=22~MeV~\cite{barn}, 
assuming that all the unusual features of the scattering may be described 
by the polarization part of the potential, particularly by the DR part.  
The parameters used for $V_{0}(r)$ are $V_{0}$=100.4~MeV, 
$r_{0}$=1.106~fm, and $a_{0}$=0.54~fm.  
                                   
 The unusual behavior of the elastic scattering and DR data for loosely 
bound projectiles can most dramatically be seen in plots of the ratios of 
the elastic differential cross section ($d\sigma_{E}/d\Omega$), 
and the DR cross section ($d\sigma_{D}/d\Omega$), to the 
Rutherford scattering cross section ($d\sigma_{c}/d\Omega$), {\it i.e.},
\begin{equation}
P_{i} \equiv \frac{d\sigma_{i}}{d\Omega}/\frac{d\sigma_{c}}{d\Omega}
 =(\frac{d\sigma_{i}}{d\sigma_{c}}), \;\;\; (i = E\;\; \mbox{or}\;\; D),
\end{equation}
as a function of the distance of closest approach $D$ (or the reduced distance 
$d$)~\cite{bass,sat2} that is related to the scattering angle $\theta$ by 
\begin{equation}
D=d(A_{1}^{1/3}+A_{2}^{1/3})= \frac{1}{2} D_{0} 
    \left( 1+ \frac{1}{\mbox{sin}(\theta/2)} \right)
    \;\;\; \mbox{with} \;\;\; D_{0}=\frac{Z_{1}Z_{2}e^{2}}{E}.
\end{equation}
Here $D_{0}$ is the distance of closest approach in a head-on collision 
($s$-wave). Further, $(A_{1},Z_{1})$ and $(A_{2},Z_{2})$ are the mass and 
charge of the projectile and target ions, respectively, and 
$E$ is the incident energy in the center-of-mass system. 
       
 In Fig.~1, we present such plots for two incident energies of 
$E$=18.5 and 21.9~MeV~\cite{agu2}.
As seen, $P_{E}$ is close to unity for large $d$, but starts to decrease 
at an unusually large distance of $d=$2.2~fm 
($\equiv d_{I}$, interaction distance). 
This value is much larger than the usual value of 
$d_{I} \approx $1.6~fm for normal, tightly bound projectiles. 
On the other hand, it is remarkable that the sum, $P_{E}+P_{D}$, remains 
close to unity until $d$ becomes as small as $\approx 1.7$~fm,
implying that the absorption in the elastic channel up to this distance, 
and the unusual character of the scattering data, is due to the breakup.

 Since the theoretical cross sections are not very sensitive to the real
polarization potential, we tentatively treat it 
in a rather crude way in carrying out 
$\chi^{2}$-analyses; we simply assume $V_{i}(r;E)$ has 
the same radial shape as the imaginary part $W_i (r;E)$:
$V_{i}(r;E)=V_{i}(E)(W_{i}(r;E)/W_{i}(E))$, $V_{i}(E)$ being the strength 
of the real potential. 
We then carry out $\chi^{2}$-analyses treating $W_{F}(E)$ and $r_{D}$ as 
adjustable parameters, keeping all other parameters fixed 
as $V_{F}$=3.0~MeV, $r_{F}$=1.40~fm, $a_{F}$=0.55~fm, $V_{D}$=0.25~MeV, 
$W_{D}$=0.40~MeV and $a_{D}$=1.25~fm. The necessity of varying  $a_{D}$
or $r_{D}$ as a function of $E$ has been shown in previous 
studies~\cite{agu1,mohr}, and in the present work we take  $r_{D}$ as 
a variable parameter to study as a function of $E$.
In the $\chi^2$-analyses,  
data for elastic scattering, angle-integrated total DR, and fusion cross
sections at $E=$14.3, 15.8, 17.3, 18.6, and 21.4~MeV are employed. 

 The values of $W_{F}(E)$ and $r_{D}(E)$ fixed from 
the $\chi^2$-analyses are presented in Fig.~2 by the open and 
the solid circles, respectively. 
Each set of circles can be
well represented by (in MeV and fm, respectively,
for $W_{F}(E)$ and $r_{D}(E)$)
\begin{equation}
W_{F}(E) \; = \; \left  \{ \begin{array}{lll}
                            0                &\;\; \mbox{, $E\leq$15.4}    \\
                            1.25(E-15.4)     &\;\; \mbox{, 15.4$<E\leq$18.5} \\
                            4.0              &\;\; \mbox{, 18.5$\leq E$} \\
                           \end{array}
                 \right.
\vspace{2ex}
\end{equation}
and
\begin{equation}
r_{D}(E)  \; = \; \left  \{ \begin{array}{lll}
                       1.73               &\;\; \mbox{, $E\leq$14.0}    \\
                       1.73-0.03(E-14.0)  &\;\; \mbox{, 14.0$<E\leq$21.4} \\
                       1.508              &\;\; \mbox{, 21.4$\leq E$}. \\
                           \end{array}
                 \right.
\vspace{2ex}
\end{equation}
Note that the threshold energy $E_{0}$=15.4~MeV, at which
$W_{F}(E)=0$, is set equal to that of the linear representation  
of quantity $S(E)=\sqrt{E\sigma_{F}} \propto (E-E_{0})$ discussed earlier.
Kolata {\it et al.} \cite{kol1} found the value to be 15.4~MeV, which 
is used in Eq.~(7).  
At this moment, we have no experimental information on $r_{D}$-values
below 14.0~MeV and above 21.4~MeV. Thus,
in Eq.~(8), we tentatively set $r_D(E)$ to be a constant as 1.73~fm 
for $E\leq$14.0~fm and 1.508~fm for $E\geq$21.4~MeV.  
Note that the values of $r_{D}(E)$ at $E$=18.6 and 21.4~MeV agree well 
with those determined by Mohr~\cite{mohr}. 

 Eqs. (7) and (8), together with other parameters used for $W_{F}(r;E)$ and 
$W_{D}(r;E)$ as mentioned above, completely fix their values 
in the energy range between $E$=14.0 and 21.4~MeV. 
In order to display the energy dependence of the potentials, 
we present in the lower panel of Fig.~3 the values of $W_{F}(r;E)$, $W_{D}(r;E)$,
and the sum $W(r;E)=W_{F}(r;E)+W_{D}(r;E)$ at a strong absorption radius 
$r=R_{sa}=13.0$~fm. It is remarkable that $W_{F}(R_{sa};E)$ 
plotted by the dotted line exhibits a threshold anomaly (strong energy variation) 
similar to that observed for tightly bound projectiles. 
However, the $W_{D}(R_{sa};E)$-values are rather flat as a function of $E$ 
and have a magnitude much larger (by about a factor of 5)
than the values of $W_{F}(R_{sa};E)$. 
Therefore, the threshold anomaly in $W_{F}(R_{sa};E)$ does not manifest itself 
in the total $W(R_{sa};E)$ plotted by the solid line.
  
 In order to generate the real part of the polarization potential by
using dispersion relations, we need to know the imaginary potential
in the entire range of $E$. Eq.~(7) with $a_{F}$=0.55~fm and $r_{F}$=1.40~fm
is enough for calculating $W_{F}(r;E)$ in the entire $E$-range. 
For the fusion potential, since the geometrical parameters are 
energy-independent, the dispersion relation is reduced to that for the 
strength parameters $V_{F}(E)$ and $W_{F}(E)$, and the closed form for the 
expression has already been obtained~\cite{maha} as
\begin{eqnarray}
V_{F}(E)=V_{s}(E_{s})
 &+&\frac{1}{\pi}W_{F}(E_{b})
       [\epsilon_{b}ln|\epsilon_{b}|-\epsilon_{a}ln|\epsilon_{a}|],
\end{eqnarray}
where 
\begin{eqnarray}
\epsilon_{a}=\frac{(E-E_{a})}{(E_{b}-E_{a})} \;\;  \mbox{and} \;\;
\epsilon_{b}=\frac{(E-E_{b})}{(E_{b}-E_{a})} \;\;
\end{eqnarray}
with $E_{a}$=15.4~MeV and $E_{b}$=18.5~MeV.  
The value $V_{s}(E_{s})$ chosen is 3.0 MeV at $E_{s}$=18.5 MeV.

 For $W_{D}(r;E)$, some care must be taken with the magnitude.
To do the initial $\chi^2$-analyses, use was made of $W_{D}$=0.4~MeV 
with $a_{D}$=1.25~fm in fixing the $r_{D}(E)$-values given by Eq.~(8). 
The constant value of $W_{D}(E)$=0.4~MeV, however, cannot
be used at very low energies, since the DR cross sections are 
expected to be extremely small in that energy region. 
The systematics of the data suggest that $\sigma_{D}$ may become essentially 
zero for $E\leq$10~MeV. 
We thus assume that  $W_{D}(E)$ increases linearly from zero at 10 MeV to 
the value of 0.4~MeV at $E$=14.0~MeV. 
The strength $W_{D}(E)$ and the radius $r_{D}(E)$ parameters 
in the entire energy range $E$ can then be rewritten as
\begin{equation}
W_{D}(E) = \left\{ \begin{array}{l}
0.0          \\
0.1(E-10.0)  \\
0.40         \\
0.40    \end{array} , 
          \right.  \;\;      
  r_{D}(E) = \left\{   \begin{array}{l}
 1.730 \hspace{3.5 cm} \mbox{,} \;\; E \leq 10.0,  \\
 1.730 \hspace{3.5 cm} \mbox{,} \;\; 10.0 \leq E \leq 14.0,\\ 
 1.730-0.03(E-14.0)\;\;\; \mbox{,} \;\; 14.0 \leq E \leq 21.4, \\
 1.508 \hspace{3.5 cm} \mbox{,} \;\; 21.4 \leq E.          \\
                                      \end{array}     
          \right.
\vspace{2ex}
\end{equation}
Together with $a_{D}$=1.25~fm, Eq.~(11) now defines $W_{D}(r;E)$ in the 
whole range of $E$.  

 In generating the real part of the DR potential, $V_{D}(r;E)$, by using 
the dispersion relation, we introduce an additional simplification 
of approximating the energy dependence of $W_{D}(r;E)$ between 
$E$=14.0 and 21.4~MeV, where $r_{D}(E)$ changes with $E$.
We assume a quadratic 
function of $E$ for $W_D (r;E)$ at each radial point $r$; 
$W_{D}(r;E)=a+b(E-E_{b})+c(E-E_{b})^{2}$, where $a$, $b$, and $c$ 
depend on $r$. 
We have confirmed that the approximation is accurate. 
Once this is done, the integration over $E$ involved in Eq.~(4) 
can be carried out analytically and one can get a closed form 
of $V_{D}(r;E)$,
\begin{eqnarray}
V_{D}(r;E)=V_{s}(r;E_{s})
 &+&\frac{1}{\pi}W_{D}(r;E_{b})
       [\epsilon_{b}ln|\epsilon_{b}|-\epsilon_{a}ln|\epsilon_{a}|] \nonumber \\
 &+& \frac{1}{\pi}(W_{D}(r;E_{c})-W_{D}(r;E_{b}))
       [\epsilon_{c}'ln|\epsilon_{c}'|-\epsilon_{b}'ln|\epsilon_{b}'|]
              \nonumber \\
 \frac{2}{\pi}(W_{D}(r;E_{c}) &+& W_{D}(r;E_{b})-2W_{D}(r;E_{m}))
[\epsilon_{c}'\epsilon_{b}'(ln|\epsilon_{c}'|- ln|\epsilon_{b}'|)+\epsilon_{b}'],
\end{eqnarray}
where $\epsilon_{a}$ and $\epsilon_{b}$ are the same as defined in Eq.~(10)
and
\begin{eqnarray}
\epsilon_{b}'=\frac{(E-E_{b})}{(E_{c}-E_{b})}, \;\; \mbox{and} \;\; 
\epsilon_{c}'=\frac{(E-E_{c})}{(E_{c}-E_{b})},
\end{eqnarray}
with $E_{a}$=10.0~MeV, $E_{b}$=14.0~MeV , $E_{c}$=21.4~MeV, and 
$E_{m}=(E_{b}+E_{c})/2$. 

 Using the polarization potentials thus generated we perform 
the final calculations for elastic scattering, total DR and 
fusion cross sections and present the results in Figs.~4 and 5, 
in comparison with the data.  
The data are fairly well reproduced by the calculations. 
The final calculated cross sections are essentially the same 
as those obtained in the initial $\chi^{2}$-analysis, showing that
the calculated cross sections do not sensitively depend on the real 
polarization potential, as we assumed in carrying out the $\chi^{2}$ analysis.
We note that the fits to the elastic scattering and reaction cross sections
(sum of the DR and fusion cross sections) are essentially the same as 
those obtained in Ref.~\cite{agu1}. 
The fit to the elastic scattering data at the lowest energy $E$=14.3~MeV is 
the worst among those shown in Fig.~4, but can be improved 
if we carry out a $\chi^{2}$-analysis including only the 
elastic scattering data as the data to be reproduced. 
We made such an analysis, finding that the data were very well reproduced 
with $r_{D}$=1.93~fm, 
much larger than $r_{D}$=1.72~fm obtained earlier. 
The DR cross section calculated with this $r_{D}$=1.93~fm, however, 
turned out to be  $\sigma_{D}$=540~mb, about 3 times larger than 
the experimental value. 
This implies that one cannot improve the simultaneous fit to both 
the elastic and DR data any further. 
                     
 In summary, we have carried out simultaneous $\chi^{2}$-analyses  
of elastic scattering, DR (breakup), and fusion cross sections for the 
$^{6}$He+$^{209}$Bi system at near-Coulomb-barrier energies within the 
framework of an optical model that introduces two types of imaginary
potentials, for DR and fusion, 
and determined the parameters of these potentials.
The results indicate that the fusion potential exhibits a threshold 
anomaly very similar to that observed for tightly bound projectiles,
but the magnitude at around the strong absorption radius is much smaller
than the imaginary DR potential that does not show such an anomaly.
Therefore, the resulting total imaginary potential does not show  
the anomaly.

 The authors sincerely thank Prof. J. J. Kolata for his kindly sending 
numerical tables of the data his group took.  
The authors also wish to express their sincere thanks to Prof. W. R. Coker
for his kindly reading the manuscript and comments.
One of the authors (BTK) acknowledges the support by Korea Research 
Foundation (KRF-2000-DP0085).

\newpage

\setlength{\leftmargin}{6em}
\begin {center}
FIGURE CAPTIONS
\end {center}
 
\vspace {1ex}\hfill\break
\par
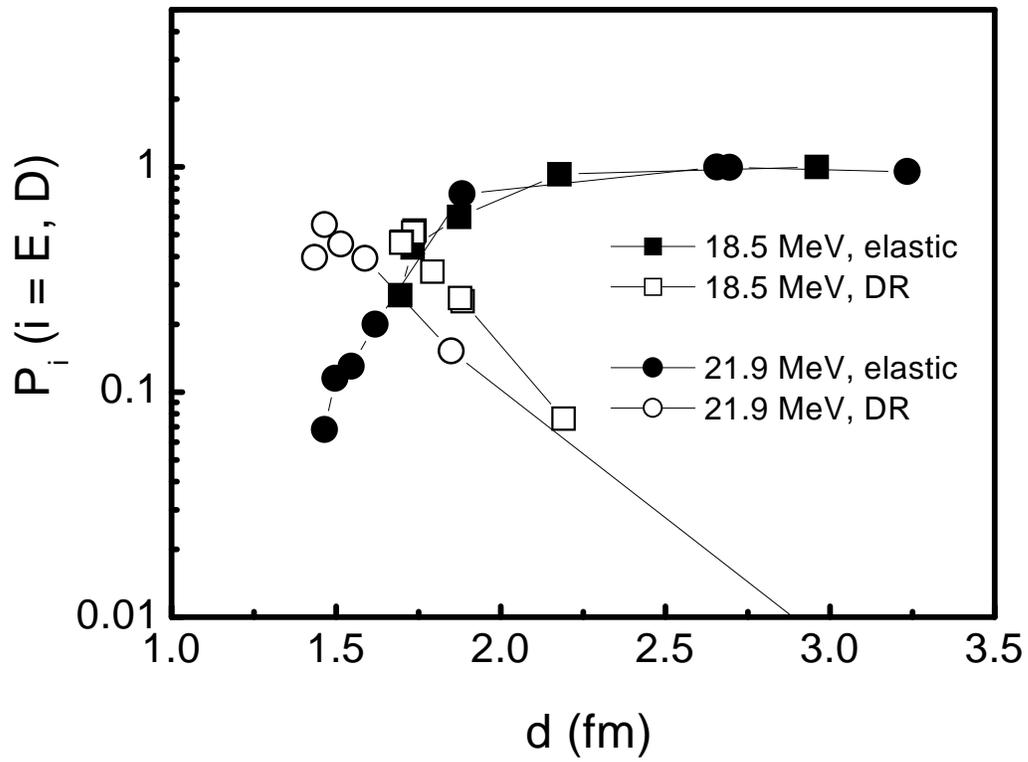
Fig.~1.~The experimental elastic and DR probabilities, $P_{E}$ and $P_{D}$, 
respectively, as a function of the reduced distance $d$ 
%for the \raisebox{1ex}{6}He+\raisebox{1ex}{209}Bi system 
for the $^{6}$He$+ ^{209}$Bi system 
at $E_{cm}=$~18.5 and 21.9 MeV. 
The data are taken from Ref.~9. The thin lines connecting
$P_i$ ($i = E$ and $D$) values are only to guide the eye.

\vspace {1ex}\hfill\break
\par
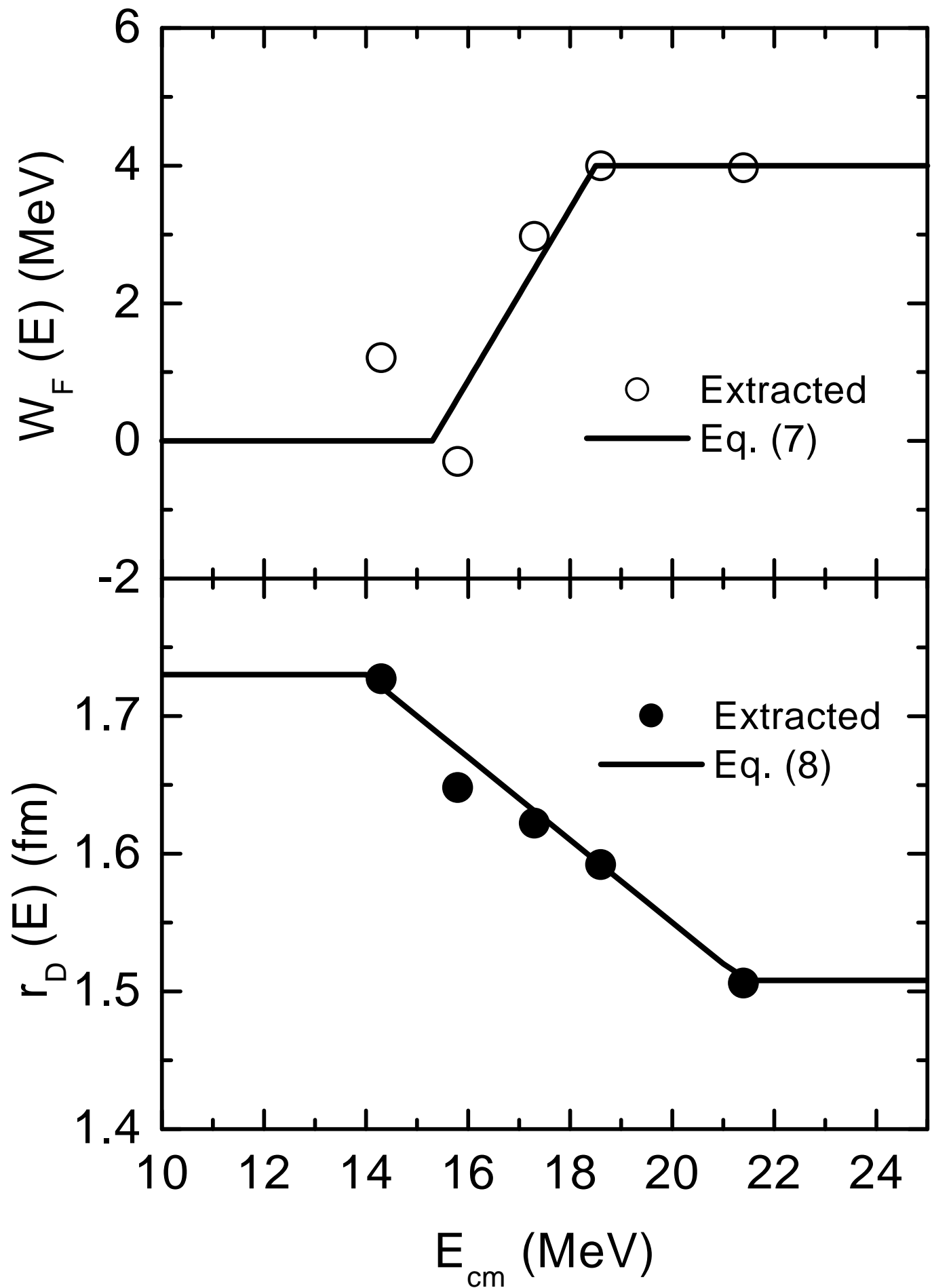
Fig.~2.~The values of $W_{F}(E)$ (upper panel) and $r_{D}(E)$ (lower panel)
extracted from the $\chi^{2}$-analyses are plotted by the open and the solid
circles, respectively. 
The solid lines represent  Eqs.~(7) and (8).

\vspace {1ex}\hfill\break
\par
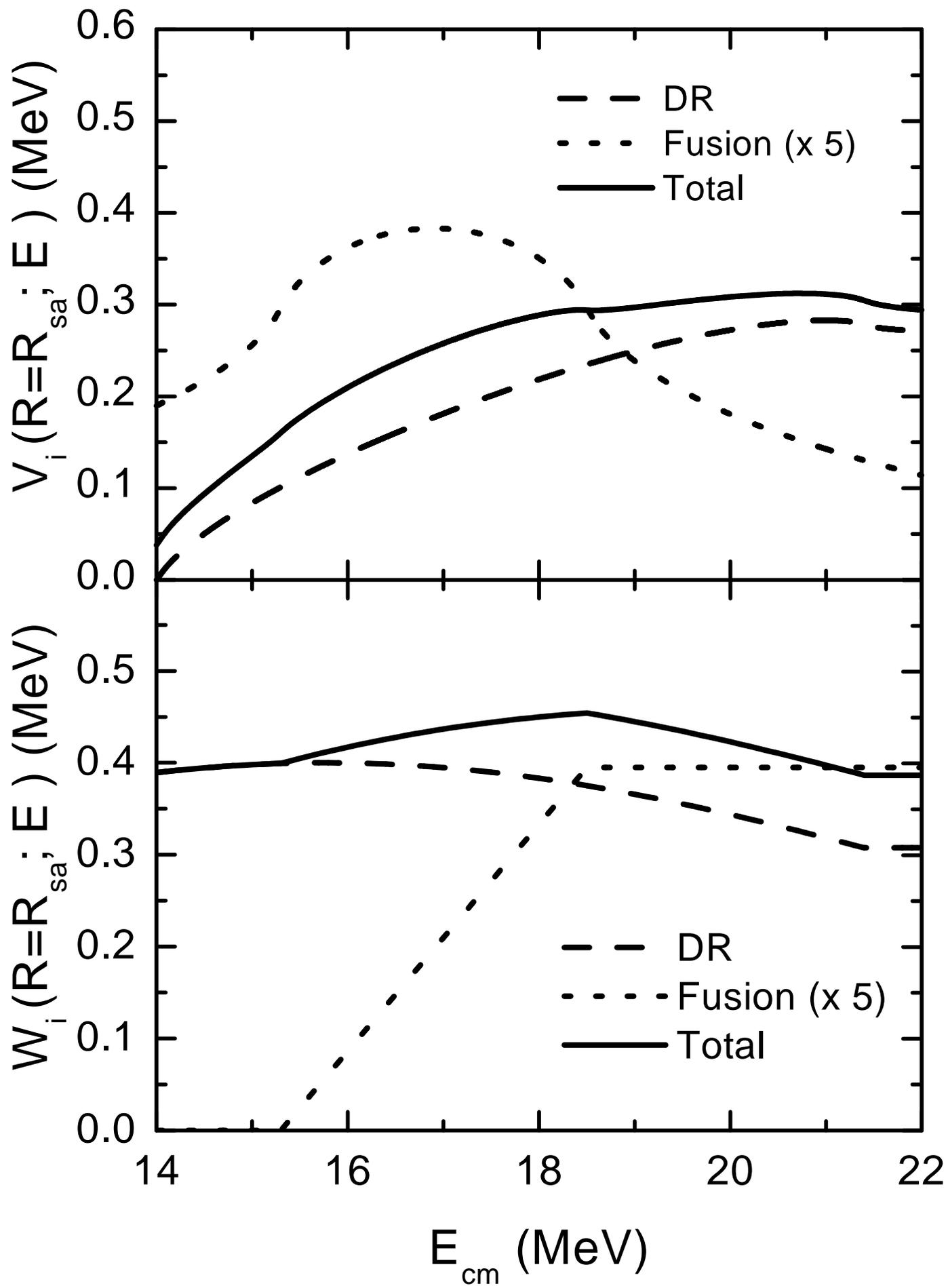
Fig.~3.~The real (upper panel) and the imaginary (lower panel) parts 
of fusion (dotted line) and DR (dashed line) potentials as functions of $E$
at the strong absorption radius $r = R_{sa} = $13.0 fm. 
The sum of fusion and DR potentials is plotted by the solid lines.
The real parts of the potentials are calculated from 
Eq.~(9) for $V_{F}(E)$, and Eq.~(12) for $V_{D}(E)$.

\vspace {1ex}\hfill\break
\par
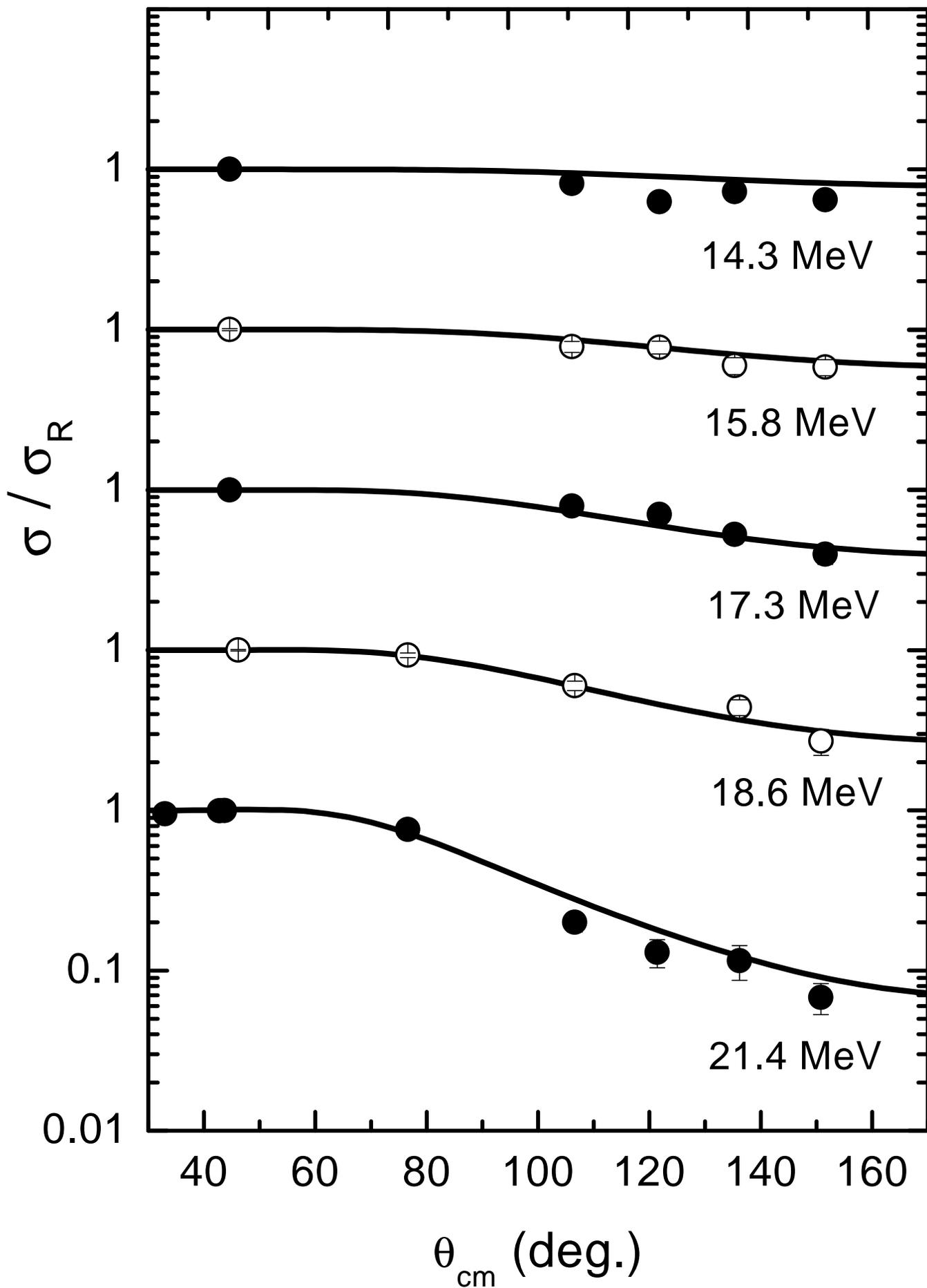
Fig.~4.~The ratios of the elastic scattering cross sections 
to Rutherford cross sections, calculated with our final optical 
%potential for the \raisebox{1ex}{6}He+\raisebox{1ex}{209}Bi system 
potential for the $^{6}$He$+ ^{209}$Bi system 
in comparison with the experimental data. 
The data are taken from Ref.~3.   

\vspace {1ex}\hfill\break
\par
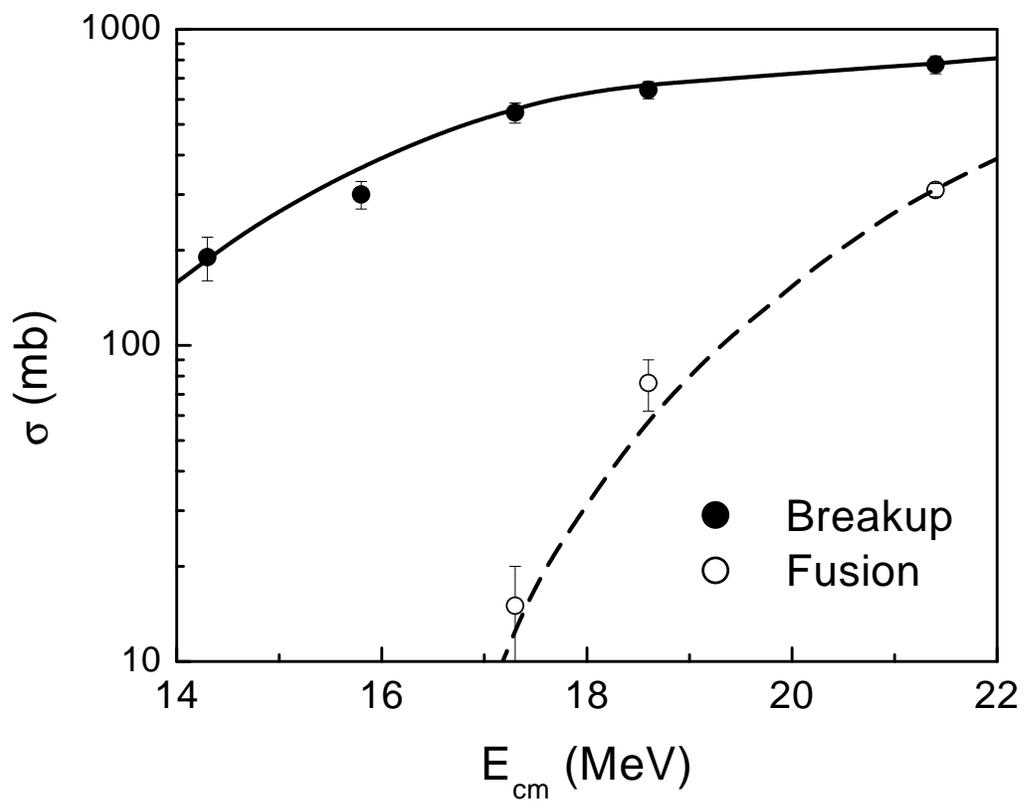
Fig.~5.~The calculated direct reaction and fusion cross sections 
with our final optical potential for the
$^{6}$He$+ ^{209}$Bi system
%\raisebox{1ex}{6}He+\raisebox{1ex}{209}Bi system
in comparison with the experimental data. 
The data are taken from Refs.~3 and 15.


\newpage

\begin{references}

\baselineskip=22pt

\bibitem{maha} C. C. Mahaux, H. Ngo, and G. R. Satchler, Nucl. Phys. 
               {\bf A449}, 354 (1986); Nucl. Phys. {\bf A456},
               134 (1986).
\bibitem{naga} M. A. Nagarajan, C. C. Mahaux, and G. R. Satchler, 
               Phys. Rev. Lett. {\bf 54}, 1136 (1985).
\bibitem{agu1} E. F. Aguilera {\it et al.}, Phys. Rev. C {\bf 63},
               061603(R) (2001).
\bibitem{tied} M. A. Tiede, D. E. Trcka, and K. W. Kemper, Phys. Rev. C
               {\bf 44}, 1698 (1991).
\bibitem{keel} N. Keeley, S. J. Bennett, N. M. Clarke, B. R. Fulton, 
               G. Tungate, P. V. Drumm, M. A. Nagarajan, and J. S. Lilly,
               Nucl. Phys. {\bf A571}, 326 (1994).
\bibitem{maci} A. M. M. Maciel {\it et al.}, Phys. Rev. C {\bf 59}, 2103
               (1999).
\bibitem{mora} S. B. Moraes {\it et al.}, Phys. Rev. C {\bf 61}, 064608 (2000).
\bibitem{sig1} C. Signorini {\it et al.}, Phys. Rev. C {\bf 61}, 061603 (2000).
\bibitem{agu2} E. F. Aguilera {\it et al.}, Phys. Rev. Lett. {\bf 84}, 5058
               (2000).
\bibitem{kell} G. R. Kelly {\it et al.}, Phys. Rev. C {\bf 63}, 024601 (2001).
\bibitem{sig2} C. Signorini {\it et al.}, in {\it Proceedings of the International
               Conference BO2000, Bologna, Italy, 2000.}, edited by D. Vretenar
               {\it et al.}, (World Scientific, Singapore, {\it in press})
\bibitem{uda1} T. Udagawa, M. Naito, and B. T. Kim, Phys. Rev. C {\bf 45},
               876 (1992).
\bibitem{stel} P. H. Stelson, Phys. Lett. B {\bf 205}, 190 (1988);
               P. H. Stelson, H. J. Kim, M. Beckerman, D. Shapira, and
               R. L. Robinson, Phys. Rev. C {\bf 41}, 1584 (1990) 
\bibitem{kim1} B. T. Kim, M. Naito, and T. Udagawa, Phys. Lett. B
               {\bf 237}, 19 (1990).                                       
\bibitem{kol1} Kolata {\it et al}., Phys. Rev. Lett. {\bf 81}, 4580 (1998).
\bibitem{mohr} P. Mohr, Phys. Rev. C {\bf 62}, 061601(R) (2000).
\bibitem{sat1} G. R. Satchler and W. G. Love, Phys. Rep. {\bf 55}, 183 (1979).
\bibitem{barn} A. R. Barnett and J. S. Lilly, Phys. Rev. C {\bf 9}, 2010
               (1974).
\bibitem{bass} R. Bass, {\it Nuclear Reactions with Heavy Ions}, 
               (Springer-Verlag, New York, 1980). 
\bibitem{sat2} G. R. Satchler, {\it Introduction to Nuclear Reactions},
               (John Wiley \& Sons, New York, 1980) p.41.

\end{references}
\end{document}